\documentclass[floats,twocolumn,amsfonts,amssymb]{revtex4}
\input epsf
\usepackage{bm}
\usepackage{amsmath}
\usepackage{amsfonts}
\usepackage{amsthm}

\newcommand{\beq}{\begin{equation}}
\newcommand{\eeq}{\end{equation}}
\newcommand{\beqa}{\begin{eqnarray}}
\newcommand{\eeqa}{\end{eqnarray}}
\newcommand{\beqan}{\begin{eqnarray*}}
\newcommand{\eenan}{\end{eqnarray*}}

\def\Bbb{\mathbb}

\newcommand{\Dslash}{{\slash{\kern -0.5em}\partial}}
\newcommand{\Aslash}{{\slash{\kern -0.5em}A}}

\def\sqr#1#2{{\vcenter{\hrule height.#2pt
     \hbox{\vrule width.#2pt height#1pt \kern#1pt
        \vrule width.#2pt}
     \hrule height.#2pt}}}

\def\thinspace{\kern .16667em}

\def\xp{x_{{\kern -.2em}_\perp}}
\def\subp{_{{\kern -.2em}_\perp}}

\def\defeq{:{\kern -0.5em}=}
\def\Tr{{\rm Tr}\,}

\def\slm{{\mathcal S}}
\def\slbar{\underline{\mathcal S}}

\def\FM{F}
\def\FL{F_{{\rm L}}}

\def\PD{{\mathcal X}_N^+}
\def\P0{{\mathcal X}_N^{++}}
\def\Nt{N}

\def\MS{{\mathcal L}_1^{+}({\mathcal H})}

\def\Trnorm#1{ \|{#1}\|_{tr} }

\newtheorem{lem}{Lemma}[section]
\newtheorem{thm}{Theorem}[section]
\newtheorem{prop}{Proposition}[section]
\begin{document}

\title{Coarse-Grained V-Representability}
\author{Paul~E.~Lammert}
\affiliation{
Dept. of Physics, 104B Davey Lab \\
Pennsylvania State University \\
University Park, PA 16802-6300}
\date{13 July, 2006} 

\begin{abstract}
The unsolved problem of determining which densities are ground 
state densities of an interacting electron system in some external
potential is important to the foundations of density functional theory. 
A coarse-grained version of this ensemble V-representability problem is 
shown to be thoroughly tractable.
Averaging the density of an interacting electron system over
the cells of a regular partition of space produces a coarse-grained 
density.  It is proved that {\em every\/} strictly positive 
coarse-grained density is coarse-grained ensemble V-representable: 
there is a unique potential, constant over each cell of the partition, 
which has a ground state with the prescribed coarse-grained density.
For a system confined to a box, the (coarse-grained) Lieb 
functional is also shown to be G\^ateaux differentiable.
All results extend to open systems.
\end{abstract}
\maketitle

\section{Introduction}
\label{intro}

Four decades ago, Hohenberg and Kohn launched modern density functional 
theory with their famous theorem\cite{HK} 
stating that a density function for a system of interacting particles 
can be a ground state density of at most one external potential.
A density is said to be {\it ensemble V-representable} (EV-representable)
if there is a mixed ground state ({\it i.e.}, a density matrix) for some 
potential which has that density.
The V-representability problem, of determining which densities are
EV-representable not only remains unresolved to this day, but has
seen almost no progress. 

Optimistically,
one might have hoped that any density which can arise from
a state with finite energy is EV-representable.  But there are known 
counterexamples\cite{ee,ccr} based on nonanalyticity.
It is also easy to construct densities, by modifying the short wavelength 
parts, which can come only from states with infinite kinetic energy, and thus 
certainly cannot be ground state densities.
These observations point to the potential usefulness of 
suppressing short length-scale degrees of freedom, or otherwise
rendering them innocuous.  Formulating the theory on a lattice 
achieves that by eliminating short wavelengths.
The lattice program was initiated by Kohn\cite{kohn} and Chayes, 
Chayes and Ruskai\cite{ccr} later proved that all densities on 
a lattice are ensemble V-representable\cite{levy,lieb,vL}.

Unfortunately, working on a lattice fundamentally alters the 
problem, and it is difficult to relate the results
to standard continuum quantum mechanics.
In this paper, I propose a way to make the short length-scale 
degrees of freedom harmless without altering the underlying quantum 
mechanics. 
Instead, by {\it coarse-graining\/}, we impose a bound on the precision 
with which we permit ourselves to specify a desired density.
Specifically, we divide our space, either all of ${\Bbb R}^3$ or
a parallelepiped, into a regular array of parellelepiped cells of 
volume $\Omega$.  The cells are labelled $B_1,B_2,\ldots$ 
(a countable infinity), or $B_1,\ldots,B_{M}$, according to the case.
Denoting the operator corresponding to the number of 
particles in cell $B_k$ by $\hat{N}_k$, we get a 
coarse-grained density operator  $\hat{{\bm \rho}}$,
with components $\hat{\rho}_k = \hat{N}_k/\Omega$.

Now we can ask about states of our system of particles which
have a specified coarse-grained density $\bm \rho$.
In particular, we want to know whether there is a potential
which has a (mixed) ground state with coarse-grained density $\bm \rho$.
If so, ${\bm \rho}$ is said to be {\it coarse-grained EV-representable}
(cg-EV-representable).
As will be shown, {\em all} coarse-grained densities, except those which 
put zero particles in one or more cells, are coarse-grained EV-representable.
Moreover, among the potentials which are constant on each cell, there
is one and only one having a ground-state coarse-grained density ${\bm \rho}$.
The proof of coarse-grained EV-representability is the central result of this
paper.  It should be clear, but bears emphasizing, that among all 
the states yielding the coarse-grained density ${\bm \rho}$, the one which is
guaranteed to be V-representable by this result minimizes 
the kinetic-plus-interaction energy.
In the case of a system of particles confined to a box, 
we show that $\FL$, viewed as a function on the affine 
hull of the set where it is finite, is G\^ateaux differentiable.  
In what sense $F_L$ can be differentiable at EV-representable
densities is a question of continuing interest\cite{swedes}.
In the unconfined case, $\FL$ is not G\^ateaux differentiable
anywhere, at least in the ordinary sense.

Coarse-graining allows short length-scale degrees of freedom to do as they
like, without artifically removing them.  Arguably, it is also more 
harmonious with practical computational limitations than the traditional 
approach.

In the standard (fine-grained) theory\cite{lieb}, it is known 
that the densities which are EV-representable are dense in the
effective domain of $\FL$ (see Thm. 3.13 of Lieb\cite{lieb} or 
Sec. I.6 of Ekeland and T\'emam\cite{ET}).
The usefulness of this observation is limited by the absence of 
any criterion to pick out the good densities, and the fact
(relevant for Kohn-Sham theory\cite{kohn-sham}) that they may depend 
on the particle interaction.  Within its scope, the coarse-grained theory 
does not suffer from these problems.
One might suspect that the coarse-grained EV-representability 
statements follow trivially from the existence of that dense set 
of densities, but it is not so because the set of fine-grained densities 
belonging to a given coarse-grained density does not contain any 
$L^1$ neighborhood.

The body of this paper unfolds as follows.
Section \ref{sketch} describes, in a rough way, 
the main ideas involved and the problems to be grappled with.
Details of the confined case (in a box) are given in Section \ref{box}.
Technically, this case is much easier than the unconfined case. 
Thus, although it stands on its own, it also serves as a warm-up 
exercise for the unconfined case, which is treated in Section \ref{nobox}.
In Section \ref{fock}, the extension to open systems with indefinite
total particle number is discussed, in somewhat less detail.  
All the results carry over, and the requisite modifications of 
the proofs are sketched.
Finally, in section \ref{conclusion}, the results are put into
perspective and directions for future work indicated.

In the body of this paper, the word {\it density} without modifier will 
carry the usual meaning, namely, ``fine-grained density.''  
Frequently, {\it coarse-grained} will be abbreviated as {\it cg}.
Coarse-grained density and number are usually expectation values;
when used in their operator aspect, a caret will be added ($\hat{N}_i$ etc.)
to help avoid confusion.  This practice is {\em not} extended to other
operators, for which ambiguity is not a danger.

\section{main ideas}
\label{sketch}

This section is an impressionistic sketch of the main ideas 
involved in the technical work of the following three sections,
concentrating on the unconfined case with fixed particle number.
As described in the introduction, we cover ${\Bbb R}^3$ with 
identical cells.  All (fine-grained) densities with the same
average density in each cell ($\rho_i$ in $B_i$) are gathered 
together into an equivalence class called a coarse-grained density.  
A cg-density ${\bm \rho} = (\rho_1,\rho_2,\ldots)$ satisfies 
$\rho_i \ge 0$ and $\sum_i \rho_i = N$.  
Dual to the cg-densities, we introduce a space ${\mathcal V}$ of 
external potentials which are constant on each cell;
other restrictions will be imposed later.
For such a potential, the integral $\int v \rho \, dx$ depends
only on the coarse-grained class of the density $\rho$.

The Lieb functional is a functional of coarse-grained density which 
represents 
the miminal {\it intrinsic energy} (kinetic plus Coulomb interaction)
consistent with the specified cg-density: 
\beq
\FL({\bm \rho}) = \inf \left\{
{\Tr} \Gamma H_0 : \, \Gamma \mapsto {\bm \rho} \right\},
\nonumber
\eeq
where $H_0$ is the kinetic-plus-interaction Hamiltonian and
the customary notation $\Gamma \mapsto {\bm \rho}$ 
indicates that mixed state $\Gamma$ has cg-density ${\bm \rho}$.
$\FL$ is not defined as a minimum because it is not obvious that 
there is any state which realizes the infimum; finite-dimensional 
intuition is a bad guide here. 
But only a state with minimal intrinsic energy in its class
can possibly be a ground state for a potential from ${\mathcal V}$,
so it is crucial for cg-EV-representability that we show $\FL$ 
really is a minimum.

To illustrate the kind of problem which can arise in 
ensuring the existence of a minimum in infinite-dimensional situations, 
consider trying to minimize the function 
$f(\psi) = \sum_{n=1}^\infty (1/n)|\langle\psi|\phi_n\rangle|^2$
over the unit vectors in a Hilbert space, where $\{\phi_n\}$ is an
orthonormal basis.  No matter how small $\epsilon > 0$, 
$S_\epsilon = \{f < \epsilon \}$ is infinite-dimensional
(and yet $\cap S_\epsilon$ is empty).
Absent detailed knowledge about the intrinsic energy,
the way to show that the {\it Levy search}\cite{levy}
implicit in the definition of $\FL$ has
a solution is to show that the search is ``almost 
finite-dimensional.''  For $\FL$, the gist of the argument
may be phrased semiclassically: 
Rough knowledge about localization of the density approximately 
confines the search to a bounded region of configuration space, 
while considering only states with relatively low intrinsic 
energy gives localization in momentum.  The two together approximately 
confine the search to a
bounded volume of phase space.  Since a volume $(2\pi \hbar)^{3N}$
of phase space corresponds to one dimension in Hilbert space,
the search is almost-finite-dimensional.

The ground state energy $E({\bm v})$ of ${\bm v} \in {\mathcal V}$ 
is the lower bound on energies of states in the presence of ${\bm v}$,
which can be expressed as
\beq
E({\bm v}) 
 =  \inf_{{\bm \rho}} \left\{ \FL({\bm \rho}) + \int {v}{\rho} \, dx \right\}.
\nonumber
\eeq
(Note that, despite the name, $E({\bm v})$ may fail to be attained,
as for the potential which is everywhere zero.)
This equation shows that $E$ is the Legendre transform of 
$\FL$.  It also shows that, for given ${\bm \rho}$, 
\beq
F({\bm \rho}) + \int {v}{\rho}\, dx \ge  E({\bm v})
\label{fve}
\eeq
holds for all ${\bm v}$.
Now, if the cg-density ${\bm \rho}$ is cg-EV-representable,
then this inequality must be an equality for the realizing potential,
and conversely, if it is equality for {\em some} ${\bm v}$, 
then ${\bm \rho}$ is clearly cg-EV-representable.

So, the task is to show that equality is achieved in
inequality (\ref{fve}).  To that end, more structure is useful. 
We embed the cg-densities in the vector space $X=\ell^1$ of real 
sequences ${\bm\rho}$ such that $\|{\bm\rho}\| = \sum |\rho_i| < \infty$, 
which is a Banach space equipped with the norm $\|\rho\|$.
Then we declare that we consider only potentials in $X^*$, 
the topological dual of $X$, which consists of ${\bm v}$ which are
{\em bounded}.  
The double Legendre transform of $\FL$ is its lower semicontinuous
convex envelope.  Therefore, by showing that $\FL$ is lower semicontinuous
and convex, we establish that $\FL$ is (up to signs) the Legendre 
transform of $E$:
\beq
F({\bm \rho}) 
 =  \sup_{{\bm v}}  \left\{ E({\bm v}) - \int {v}{\rho}\, dx.\right\}.
\nonumber
\eeq
Convexity is immediate from the definition, but lower semicontinuity
requires some work. 

In the case of a system confined to a box, $X$ and $X^*$ are
finite-dimensional.  In that case, existence of a potential which 
gives equality in Eq. (\ref{fve}) is an automatic consequence of
the mutual Legendre transform relation of $E$ and $\FL$.  

The unconfined case is harder.  We attack it by noting that
the second Legendre transform relation says that equality in 
Eq. (\ref{fve}) can be approached as closely as desired.  
That is, a sequence of ${\bm v}\in X^*$ can be found such that 
\beq
E({\bm v}_n) - \int {v}_n{\rho}\, dx \to F({\bm \rho}).
\nonumber
\eeq
The trick is then to show that this sequence ${\bm v}_n$
converges in some sense and that the limit ${\bm v}_\infty$
satisfies equality in (\ref{fve}).  
In carrying that out, it is necessary, in general, to locate 
the limit outside of $X^*$ entirely.    
This may seem surprising, at first, but was entirely to be
expected.  After all, the harmonic oscillator ground state
is v-representable, but the harmonic-oscillator potential is
not bounded, hence not in $X^*$.

This, in brief, is how cg-EV-representability is established.
We also prove a coarse-grained version of the Hohenberg-Kohn 
Theorem, with only a little more work than for the usual version.  
With that, we can see that, not only is every everywhere-positive 
${\bm\rho}$  cg-EV-representable, but that the representing 
potential is unique.

\section{confined systems}
\label{box}

In this section, we treat confined systems.  Standard tools of convex 
analysis will be used, without much comment apart from reminders of 
definitions.  The reader for whom the material is unfamiliar can 
find it in a number of textbooks, for example, chapter I of Ekeland and 
T\'emam\cite{ET}, or van Tiel\cite{vT}.
Rockafellar\cite{rockafellar} treats the finite-dimensional theory
exhaustively.  In citing specific results in those works, the
abbreviations ET, vT and R are used. 

We are concerned with a system of $\Nt$ fermions (``electrons,'' with
or without spin) 
in a box, which we take to be a parallelepiped for simplicity.  
The Hilbert space of antisymmetrized states for these
particles is denoted $\mathcal H$. 
A mixed state, not necessarily normalized, has the form
$\Gamma = \sum_{n=1}^\infty \alpha_n | \psi_n\rangle\langle \psi_n|$,
where $\psi_n$ are normalized states in $\mathcal H$,
$0 \le \alpha_n$ for each $n$, and 
$\sum_{n=1}^\infty \alpha_n < \infty$.
The set of normalized mixed states, which are those satisfying
$\Tr \Gamma = \sum_{n=1}^\infty \alpha_n = 1$, will be denoted
simply by $\slm$.
The mixed states belong to the space ${\mathcal L}_1({\mathcal H})$ 
of trace class operators on ${\mathcal H}$.
More specifically $\slm$ is the intersection of the convex cone 
$\MS$ of positive operators with the hyperplane $\Tr \Gamma = 1$.
${\mathcal L}_1({\mathcal H})$ is a Banach space under the
trace norm $\Trnorm{A} = \Tr |A|$, where $|A|$ is the unique 
positive square root of $A^\dagger A$.

The box is subdivided into a regular array of parallelepiped cells
$B_1,\cdots,B_M$ of volume $\Omega$, with associated density operators
$\hat{N}_1/\Omega,\cdots,\hat{N}_M/\Omega$.  Generally, we look upon 
sequences of this sort as points in ${\Bbb R}^{M}$, and denote
them by bold-face symbols.
$\slm$ decomposes into strata labeled by 
${\bm \rho} \in {\Bbb R}^{M}$: 
\beq
{\slm}({\bm \rho}) = \{ \Gamma \in \slm : \,
{\Tr} \Gamma \hat{\bm N} = {\bm \rho} \}.
\eeq
The space into which cg-densities have been embedded is generically
denoted by ${\mathcal X}$.  In this section, ${\mathcal X} = {\Bbb R}^M$.
The term {\it cg-density} is reserved for elements of ${\mathcal X}$
which have only nonnegative components, denoted by ${\mathcal X}^+$.  
The set of {\em normalized\/} cg-densities is
\beq
{\PD} = \left\{ {\bm \rho}\in {\Bbb R}^{M}:
{\bm \rho} \ge 0, \,\,
\sum_{i=1}^M \rho_i \Omega = \Nt
\right\}.
\eeq
If ${\bm\rho}$ is not a normalized cg-density, 
then ${\slm}({\bm \rho})$ is empty.
${\PD}$ is an $(M-1)$-dimensional simplex contained in the 
$M-1$ dimensional hyperplane ${\mathcal X}_N$ of 
${\bm \rho}\in {\Bbb R}^M$ such that $\Omega \sum \rho_i = N$.
The interior of ${\mathcal X}_N^+$ relative to ${\mathcal X}_N$
corresponds to normalized cg-densities which are {\em greater} than
zero in every cell, and will be denoted by $\P0$.
(Standard nomenclature\cite{rockafellar} calls ${\mathcal X}_N$ 
the {\it affine hull} of $\PD$, and $\P0$ the {\it relative interior} 
of $\PD$.)

The Hamiltonian $H_0$ includes kinetic energy and the interaction 
between the electrons, with periodic, Dirichlet or Neumann boundary
conditions.  We assume the interaction is the repulsive
Coulomb interaction, possibly modified by a bounded interaction,
such as would be required to adapt it to periodic boundary conditions.
The bounded addition may even be a multiparticle interaction, but we
assume it has a constant adjusted so that the total interaction is
nonnegative.  $H_0$ is then unbounded, but bounded below by zero.

There is a lower limit to the kinetic plus interaction energy of
states having a given cg-density, the Lieb functional 
\beq
\FL({\bm \rho}) \defeq \inf \left\{
{\Tr} \Gamma H_0 : \, \Gamma \in {\slm}({\bm \rho}) \right\}.
\label{FL}
\eeq
If ${\slm}({\rm \rho})$ is empty, $\FL({\bm \rho}) = \infty$.
It may appear at first sight that the trace in the definition is
potentially not well-defined, since $H_0$ is unbounded. 
However, since $H_0 > 0$ and $\Gamma$ is a {\em positive} operator, 
there is no real difficulty.  We can, for example, replace
$H_0$ by $H_0 P_{H_0}(\lambda)$, where $P_{H_0}(\lambda)$ is the 
spectral projection onto $H_0 \le \lambda$, and take the limit 
$\lambda \to \infty$. 
The set of states which realize the infimum in Eq. (\ref{FL})
is denoted ${\slbar({\bm \rho})}$:
\beq
{\slbar({\bm \rho})}
\defeq \left\{ \Gamma \in {\slm}({\bm \rho}) :
{\Tr} \Gamma H_0 = \FL({\bm \rho}) \right\}.
\eeq
We will show later that this set is nonempty.  $\FL$ is bounded both 
above and below: $0 < \FL \le {\FL}^{max}$.  The lower bound is simple,
and the upper bound follows by using the fact ($N$-representability)
that a smooth density with any desired cg-density can be realized by 
an appropriate mixed state, and then using the bounds on $\FL$ in 
terms of $\nabla \rho$ as derived in Lieb\cite{lieb} 
(see Thms. 3.8 and 3.9 there).  The largest of these (finite) bounds
is for the case of all particles in a single cell, 
so $\FL$ is bounded above. 

Having dealt with densities, we now turn to potentials.
The potentials we want to consider take the constant value
$v_i$ on cell $B_i$, information which is encoded in the vector
${\bm v} = (v_1,\ldots,v_M) \in {\Bbb R}^M$.  We will write
${\mathcal V}$ for this space of potentials.
Adding the potential ${\bm v}$ to $H_0$ gives
\beq
H({\bm v}) = H_0 + \sum_{i=1}^M \Omega v_i \hat{\rho}_i.
\eeq
When convenient, we use the alternate, and suggestive, notations
$\int v \, \hat{\rho}({\bm x}) \, d{x}$ or ${\bm v}\cdot\hat{\bm \rho}$
for the potential term.

The {\it ground state energy} for potential ${\bm v}\in {\mathcal V}$ is
\beq
E({\bm v}) 
\defeq \inf \left\{ 
{\Tr} \Gamma H({\bm v}) \, : \, \Gamma \in 
\slm \right\}.
\nonumber
\eeq
This really is the energy of some state; the infimum is sure
to be realized in the present setting, as follows from the fact 
that our system is in a box (cf. Thm. \ref{minima}).
We can rewrite $E({\bm v})$ in terms of the Levy constrained search as 
\beq
E({\bm v}) 
=  \inf_{{\bm \rho}} \{ \FL({\bm \rho}) + {\bm v}\cdot{\bm \rho} \} 
\label{GSE2}
\eeq
That is, $-E(-{\bm v})$ is the {\it Legendre-Fenchel transform}
(also known as {\it convex conjugate}, {\it conjugate}, or {\it polar}) 
of $\FL({\bm \rho})$.  
Thus, 
$E({\bm v})$ is concave and upper semicontinuous where it is finite.
But, the infimum in Eq. (\ref{GSE2}) is certain to lie between 
$\min\{v_i\}$ and $\FL^{max} + \max\{v_i\}$, so that 
$E({\bm v})$ is finite everywhere on ${\mathbb R}^M$, and therefore
continuous.
The Legendre-Fenchel transform of $-E(-{\bm v})$ is the closed
convex hull of $\FL$.  $\FL$ is easily seen to be convex from
its definition, so continuity on 
$\P0$ is 
automatic because of finite dimensionality 
(ET: Cor. I.2.3, vT: Thm. 5.23, R: Thm. 10.1), 
but lower semicontinuity at the relative boundary of ${\PD}$ is not.
In fact, $\FL$ is lower semicontinuous there as well.
Rather than provide the proof (see Thm. \ref{FL-lsc}), we will here
simply assume that $\FL$ is {\em defined} on the relative boundary 
so that it is lower semicontinuous.  
This is harmless because the cg-densities there are not of interest.
We summarize the foregoing discussion.
\begin{prop}[convex duality]
\label{duality}
The functions $\FL({\bm \rho})$ and $E({\bm v})$ on ${\Bbb R}^M$ are 
in duality, {\it i.e.,} they are (up to signs) Legendre transforms 
of each other.  Hence, $\FL$ and $-E$ are convex and lower semicontinuous.
\end{prop}
One aspect of this relationship between $\FL$ and $E$ which will
be important later is that ${\bm v}\in{\mathcal V}$ is regarded as a 
linear functional on ${\Bbb R}^M$ via ${\bm \rho}\mapsto \int v\rho \,dx$.  
Since ${\Bbb R}^M$ is finite-dimensional, ${\bm v}$ is automatically a
continuous functional (though $v(x)$ is not a continuous function of $x$
unless it is constant).

Now, we are ready for a coarse-grained version of the Hohenberg-Kohn
theorem.
If potentials ${\bm v}$ and ${\bm v^\prime}$ differ by an overall 
constant, write ${\bm v} \sim {\bm v^\prime}$, and denote
the equivalence class of ${\bm v}$ in ${\mathcal V}/{\mathbb R}$
by $[{\bm v}]$.

\begin{thm}[{\bf cg Hohenberg-Kohn}] 
\label{HK}
Given $\slm({\bm \rho})$, there is at most one equivalence class in 
${\mathcal V}/{\mathbb R}$
which has a ground state in $\slm({\bm \rho})$.
\end{thm}
\begin{proof}
First, ${\bm v} \not\sim {\bm v^\prime}$ implies that
$H({\bm v})$ and $H({\bm v^\prime})$ have disjoint ground state
manifolds in $\mathcal H$.  
This is a result about partial differential equations
and is proven just as for the conventional Hohenberg-Kohn 
theorem, being just a specialization to potentials in ${\mathcal V}$.

If ${\bm v} \in {\mathcal V}$ has a ground state in 
$\slm({\bm \rho})$, then that state is in ${\slbar({\bm \rho})}$, 
and {\em all} elements of ${\slbar({\bm \rho})}$ are ground states 
for ${\bm v}$, since all such states have the same value of
$\Tr H_0 \Gamma$.
Thus, if both ${\bm v}$ and ${\bm v^\prime}$ have ground
states in $\slm({\bm \rho})$, then they share a ground state, 
which contradicts the previous paragraph.
\end{proof}

This proof carries over to the unconfined setting of Section 
\ref{nobox} and the Fock space setting of Section \ref{fock}, 
with no essential changes.

As just observed, the only states in $\slm({\bm \rho})$ 
which can possibly be ground states are those in $\slbar({\bm \rho})$.
The next task is therefore to show that $\slbar({\bm\rho})$ is
nonempty.

\begin{thm}[{\bf Existence of Levy search solutions}]
\label{minima}
For every ${\bm \rho}\in {\PD}$, 
${\slbar({\bm \rho})} \not= \emptyset$.
That is, there is a normalized mixed state $\Gamma$ such that
$\FM({\bm \rho}) = \Tr H_0 \Gamma$. 
\end{thm}

We prove this by means of a couple of lemmas.
It will be useful to write the intrinsic energy in a
functional notation:
\beq
{\mathcal E}(\Gamma) \defeq {\Tr} \Gamma H_0,
\eeq
regarded as a function on $\slm$, so that
${\mathcal E}^{-1}(0,M]$ consists of {\em normalized}
mixed states with intrinsic energy not exceeding $M$.  

\begin{lem}
\label{E0-lsc}
${\mathcal E}$ is lower semicontinuous with respect to
trace norm on $\slm$.
\end{lem}
\begin{proof}
Suppose $\Gamma_n \in {\mathcal E}^{-1}(0,M]$ for $n=1,2,\ldots$
and $\Gamma_n {\to} \Gamma$ in trace norm.  We need to show
that ${\mathcal E}(\Gamma) \le M$.  Suppose instead that
${\mathcal E}(\Gamma) > M$.  

If $P_{H_0}(e)$ denotes the spectral projection onto $H_0 \le e$,
${\Tr} \Gamma H_0 P_{H_0}(e)$ is monotonic in $e$ so that
there is some finite $e$ for which ${\Tr} \Gamma H_0 P_{H_0}(e) > M$.
But, trace-norm convergence implies weak convergence and 
$H_0 P_{H_0}(e)$ is a bounded operator.  Therefore, 
${\Tr} \Gamma_n H_0 P_{H_0}(e)$ exceeds $M$ for large enough
$n$ since it converges to ${\Tr} \Gamma H_0 P_{H_0}(e)$.
This, however, is impossible because
$M \ge {\mathcal E}(\Gamma_n) \ge {\Tr} \Gamma_n H_0 P_{H_0}(e)$.
\end{proof}

This result is very general, depending as it does only on the 
positivity of mixed states and the semiboundedness of the Hamiltonian.
The theorem and proof apply to the unconfined case, and even
in Fock space.

For the next lemma, we recall the concept of {\it total boundedness}.
A set in a metric space is called totally bounded if, given $\epsilon > 0$,
it can be covered by a finite number of balls of radius $\epsilon$.
For a subset of a complete metric space, total boundedness is
equivalent to relative compactness, {\it i.e.,} having a compact
closure.  A Banach space is a complete metric space.

\begin{lem}
\label{E0-cpt}
${\mathcal E}^{-1}(0,M]$ for $M < \infty$ is relatively compact 
with respect to trace norm.
\end{lem}
\begin{proof}
 
${\mathcal E}^{-1}(0,M]$ is certainly bounded since it is contained 
in the unit ball of $\MS$.  The strategy is to show that, given
$\epsilon$, there is a finite dimensional space $W$, such that
all points of ${\mathcal E}^{-1}(0,M]$ are within 
distance $\epsilon$ of $W$.  Since a bounded subset of $W$
is totally bounded and $\epsilon$ is arbitrary, this will show 
that ${\mathcal E}^{-1}(0,M]$ is totally bounded, hence relatively
compact.

The interaction energy is nonnegative, so
${\mathcal E}^{-1}(0,M]$ can only get bigger if we drop the
interaction.  Thus, it suffices to prove the lemma for the 
case that $H_0 = T$ contains only kinetic energy.

Due to our assumptions, the single-particle eigenstates
$\varphi_m$ of $T$ can be written down explicitly, and 
the spectrum is certainly discrete.  
Up to energy $M/\epsilon$, there are only a finite number, 
$\varphi_1,\ldots,\varphi_K$, and the set of Slater determinants
made from these likewise span a finite-dimensional space, $W$.

Now, any state $\Gamma$ can be written as a linear combination 
of Slater determinants of the $\varphi_m$.  But, if 
${\mathcal E}(\Gamma) \le M$, the terms containing
$\varphi_m$ for $m > K$ must, all together, have norm less
than $\epsilon$.  That is, $\Gamma$ is the sum of something
in $W$ and something with norm less than $\epsilon$.
\end{proof}

The box played a crucial role in this proof; its counterpart
in the unconfined case will be much more difficult.

\begin{proof}{(of Theorem \ref{minima})}
Since the $\hat{N_i}$ are bounded operators, the 
condition ${\Tr} \Gamma\hat{{\bm N}} = {\bm N}$ defines a 
closed linear subspace of ${\mathcal L}_1({\mathcal H})$. 
The intersection of this subspace with
${\mathcal E}^{-1}(0,M]$ is therefore compact.
${\mathcal E}$ is lower semicontinuous by (Lemma \ref{E0-lsc}).
Since a lower semicontinuous function on a compact
space is guaranteed to have a minimizer, the theorem is proven.
\end{proof}

A cg-density ${\bm\rho}_0$ is cg-EV-representable if and only if some
$\Gamma\in\slbar({\bm\rho}_0)$ is a ground state of some
${\bm v}\in{\mathcal V}$. 
Since Theorem \ref{minima} shows that $\slbar({\bm\rho}_0)$ is nonempty, 
the condition for cg-EV-representability is 
existence of ${\bm v}\in{\mathcal V}$ such that
$\FL({\bm \rho}_0) + \int v \rho_0 \, dx \le
\FL({\bm \rho}) + \int v \rho \, dx$, for all ${\bm \rho}$.  
This condition rearranges to the statement that the hyperplane 
${\bm\rho} \mapsto \FL({\bm \rho}_0) + \int v (\rho - \rho_0) \, dx$
lies below the graph of $\FL$ and touches it at ${\bm \rho}_0$.
That is, ${\bm v}$ is a tangent functional to $\FL$ at ${\bm \rho}_0$
and, since it is continuous, 
a {\it subgradient\/}\cite{rockafellar,ET,vT} at ${\bm\rho}_0$.
Recall that the set of all subgradients at ${\bm \rho}$
[denoted $\partial \FL({\bm \rho})$], is called the 
{\it subdifferential\/} of $\FL$ at ${\bm \rho}$, and 
$\FL$ is said to be {\it subdifferentiable\/} at ${\bm \rho}$
if $\partial \FL({\bm \rho}) \not= \emptyset$.
It is a basic fact (ET: Prop. I.5.2, vT: Thm. 5.35) 
that a convex function on a finite-dimensional space is 
subdifferentiable everywhere on the relative interior of its 
effective domain.  This observation completes the proof of the 
central result of this section:

\begin{thm}[{\bf Ensemble $v$-representability}] 
\label{v-rep}
Every ${\bm \rho}$ in $\P0$ 
is cg-EV-representable.
\end{thm}
\begin{proof} See preceding discussion.
\end{proof}

Combining this theorem with the cg-Hohenberg-Kohn Theorem
gives G\^ateaux differentiability\cite{ET,vT,bb}
of the restriction of $\FL$ to ${{\mathcal X}_N}$.

\begin{thm}[{\bf Differentiability of $\FL|_{{\mathcal X}_N}$}]
$\FL$, as a function on ${\mathcal X}_N$,
is G\^ateaux differentiable on $\P0$. 
\end{thm}
\begin{proof}
Given ${\bm\rho}\in \P0$, 
Thm. \ref{v-rep} says that there is at least one 
${\bm v}$ in ${\mathcal V}$ which represents $\bm\rho$.  
On the other hand, Theorem \ref{HK} asserts that there is at most one.
So, there is precisely one.  Uniqueness of the subgradient at
$\bm\rho$ together with local boundedness of $\FL$ shows 
(ET: Prop. I.5.3) that
$\FL|_{{\mathcal X}_N}$ is G\^ateaux differentiable at $\rho$.
\end{proof}

Note in passing that we have no comparable theorem for $E({\bm v})$.
If $\bm v$ has distinct ground states which lie in different strata
of $\slm$, then $E$ is certainly not G\^ateaux differentiable there.

\section{unconfined systems}
\label{nobox}

Of course, the box used in the previous section is a 
rather artificial element, so it is desirable to prove analogous 
results with all space accessible to the particles.  
We refer to this as the unconfined case, and in this section
we will show cg-EV-representability of all 
normalized, strictly positive, cg-densities for the unconfined case.

Similarly to before, ${\mathbb R}^3$ is partitioned into a 
regular, countably infinite, array of cells $B_1, B_2, \ldots$, 
each of volume $\Omega$.  
This time, the regularity serves the important purpose of ensuring 
uniform upper and lower bounds on the cell sizes.
As in Section \ref{box}, 
$\slm$ denotes the set of normalized mixed states,
$\PD$ the set of normalized cg-densities, and
$\P0$ those which are nowhere zero. 
Now, however, ${\mathcal X}$ is $\ell^1$, the space of sequences 
$(\rho_1,\rho_2,\ldots)$ such that $\|{\bm\rho}\| = \sum |\rho_i| < \infty$.
This space is a Banach space when equipped with the norm
$\|{\bm \rho}\|$.
The Lieb functional is defined as before, and has the same
upper bound ${\FL}^{max}$ on $\P0$, outside of which it is $+\infty$.
$\PD$ is contained in the closed subspace 
${\mathcal X}_N = \{{\bm \rho}\in\ell^1 : \Omega\sum\rho_i = \Nt\}$;
however, the interior of $\PD$ relative to ${\mathcal X}_N$
is empty, not $\P0$.

Naturally, we then consider potentials in the dual space to $\ell^1$,
which is the space $\ell^\infty$ of sequences $(v_1,v_2,\ldots)$
which are bounded: ${\rm sup}\ |v_i| < \infty$.
However, we shall eventually find ourselves looking for
potentials in the larger set 
\beq
{\mathcal V} = \left\{{\bm v} = (v_1,v_2,\ldots)\, 
:\, \inf v_i > -\infty \right\},
\eeq
of potentials which are required only to be bounded {\em below}.
Note that the cg-Hohenberg-Kohn Theorem \ref{HK} works for potentials
in ${\mathcal V}$.

From the experience of the box case, one might expect the work
here to split into two principal parts:
proof of the existence of Levy search solutions,
by demonstration that $\FL$ is lower semicontinuous and that the
search is taking place in a compact space, 
followed by proof of subdifferentiability using general results
from the duality theory of convex analysis.
The first expectation is correct.
Lemma \ref{E0-lsc} goes through unaltered, but Lemma \ref{E0-cpt} 
will require a more sophisticated counterpart.
The run-up to Theorem \ref{minima-nobox} uses some ideas from 
Lieb\cite{lieb} (particularly Thm. 4.4 there),
but the proof given here is self-contained. 
However, what convex analysis gives us in this case is 
limited to showing that $\FL({\bm\rho})$ and $E({\bm v})$ are
a Legendre-Fenchel transform pair (up to signs).  
Establishing existence of a potential ${\bm v}$ which has
ground state with density ${\bm\rho}\in \P0$ takes an
{\it ad hoc} argument following the pattern of the proof by
Chayes, Chayes and Ruskai\cite{ccr} of EV-representability 
on a lattice.  The basic reason for this is that some of the
potentials needed do not lie in the dual space to $\ell^1$.
In fact, they are not even {\em linear functionals}, because
they take the value $+\infty$ on some densities.  

The immediate goal is the general-purpose Thm. \ref{tight-cpt} below, 
which deals with (fine-grained) densities.
Call a set $\Delta$ of densities {\it tight\/} 
if, given $\epsilon > 0$, there exists a bounded set 
$\Lambda_\epsilon$ such that no $\rho \in \Delta$ puts more than a 
total of $\epsilon$ particles outside $\Lambda_\epsilon$. 
Rephrasing, $\Delta$ is tight if, given a tolerance, there is
a phantom box which contains all the densities to that tolerance.
The set of mixed states whose densities belong to 
$\Delta$ is denoted by ${\mathcal S}({\Delta})$.
The theorem says, roughly, that a set of phantom boxes is 
enough to guarantee that the low energy states among
${\slm}({\Delta})$ are an almost finite-dimensional set,
which is one of the important services the real box of the previous 
section performed. 

Some notation that will be used is gathered here:
For $\Lambda$ a bounded region, denote the Hilbert space projection 
corresponding to {\em all\/} particles being in $\Lambda$ by
$P_\Lambda$.
$H_0 +1$ is bounded below by 1, and therefore has a unique positive 
square root denoted by $h$ which is also bounded below by 1. 
$h^{-1}$ is a bounded operator.

The next lemma is key.  It is followed by three auxiliary results,
the proofs of which can be skipped without detriment to understanding
of the main results.
\begin{lem}
\label{PDP-cpt}
For $\Lambda$ a bounded region of space, 
$P_\Lambda {\mathcal E}^{-1}(0,M] P_\Lambda$ is compact.
\end{lem}
\begin{proof}

Let $(P_\Lambda \Gamma_n P_\Lambda)_{n=1}^\infty$ be a bounded
sequence in 
$P_\Lambda {\mathcal E}^{-1}(0,M] P_\Lambda$.
We will show that there is a norm convergent subsequence.
Since $\Gamma_n \in {\mathcal E}^{-1}(0,M]$, 
$\gamma_n = h\Gamma_n h$ is also a bounded sequence in $\MS$. 
Now,  
${\mathcal L}_1({\mathcal H})$ is the topological dual to
the space of compact operators on ${\mathcal H}$, and any bounded 
sequence in ${\mathcal L}_1({\mathcal H})$, in particular ($\gamma_n$),
has a weak-$^*$ convergent subsequence.
[This is a slightly subtle point.  It requires both the Banach-Alaoglu
theorem and the fact that since ${\mathcal H}$ is separable, the
space of compact operators on ${\mathcal H}$ is also separable.
See, e.g., Thm. 3.17 of Rudin\cite{rudin}.]
By relabeling, write the weak-$*$ convergent subsequence again
as $\gamma_n$, so for any compact operator $A$,
$\Tr \gamma_n A \to \Tr \gamma A$, where $\gamma$ is the limit.
But, the Hamiltonian $H_0$ is locally compact\cite{HS},
{\it i.e.,} $P_\Lambda h^{-1}$ and $h^{-1}P_\Lambda$
are compact operators (Lemma \ref{H0-lcl-cpt}).
Since the product of a bounded operator and a compact operator
is compact, for any bounded operator $B$,
$\Tr B P_\Lambda \Gamma_n P_\Lambda  = 
\Tr \{(B P_\Lambda h^{-1}) \gamma_n (h^{-1} P_\Lambda)\}
\to \Tr B P_\Lambda h^{-1} \gamma h^{-1} P_\Lambda$.
Defining $\Gamma := h^{-1}\gamma h^{-1}$, this says
{$P_\Lambda \Gamma_n P_\Lambda  {\to} P_\Lambda \Gamma P_\Lambda$}
with respect to the weak topology (not weak-$^*$!).
Now apply Lemma \ref{wk2norm} to deduce
{$P_\Lambda \Gamma_n P_\Lambda  \stackrel{{\rm tr}}{\to}
P_\Lambda \Gamma P_\Lambda$}.
So, $P_\Lambda {\mathcal E}^{-1}(0,M] P_\Lambda$ is relatively
compact; but it is also closed, by Lemma \ref{E0-lsc}. 
\end{proof}

\begin{lem}
\label{matrix-blocks}
For $\Gamma \in \MS$ and $P$ an orthogonal projection on ${\mathcal H}$,
if $\Trnorm{P \Gamma P} > (1 -\epsilon^2/9)\Trnorm{\Gamma}$,
and $\epsilon < 3$, then
$\| \Gamma - P \Gamma P \|_{tr} < \epsilon\Trnorm{\Gamma}$.
\end{lem}
\begin{proof}
Without loss, assume $\Gamma$ is normalized, so
$\Gamma = \sum c_m |\phi_m\rangle\langle \phi_m|$,
where $c_m > 0$, $\sum c_m = 1$ and $\phi_m$ are unit vectors.
Write $\Gamma = P \Gamma P + P \Gamma P^\perp + P^\perp \Gamma P
+ P^\perp \Gamma P^\perp$, with $P^\perp = 1-P$, and bound the last 
three terms in trace norm as follows.
$$
\Trnorm{P^\perp \Gamma P^\perp} = \Tr {P^\perp \Gamma P^\perp} 
= 1 - \Tr {P \Gamma P} < \epsilon^2/9,$$ 
where the final inequality is by hypothesis.
Using the triangle inequality followed by the Cauchy-Schwarz
inequality and then the above bound on $\Trnorm{P^\perp \Gamma P^\perp}$,
we get $\Trnorm{P\Gamma P^\perp} = \Trnorm{P^\perp \Gamma P} 
\le \sum c_m \|P^\perp \phi_m\|\, \| P\phi_m\|
\le \left(\sum c_m \|P^\perp \phi_m\|^2\right)^{1/2}
\left(\sum c_m \|P \phi_m\|^2\right)^{1/2}
\le \epsilon/3$.  
Gathering the pieces, and using $\epsilon < 3$, gives the result.
\end{proof}

\begin{lem}
\label{wk2norm}
If $(\Gamma_n)_{n=1}^\infty$ is a sequence in $\MS$, then
$\Gamma_n \stackrel{wk}{\to} \Gamma$ if and only if
$\Gamma_n \stackrel{tr}{\to} \Gamma$, and in that case
$\Gamma \in \MS$.
\end{lem}
\begin{proof}
The implication from norm convergence to weak convergence is trivial, 
so suppose $\Gamma_n {\to} \Gamma$ weakly.
The cone $\MS \subset {\mathcal L}_1({\mathcal H})$ is norm closed and
convex, hence weakly closed.  Thus, any weak limit of $(\Gamma_n)_1^\infty$
is also in $\MS$.
Since $\Trnorm{\Gamma_n} = \Tr 1\Gamma_n \to \Tr \Gamma  = \Trnorm{\Gamma}$,
the case $\Gamma = 0$ is easy, and we may as well assume the $\Gamma_n$
and $\Gamma$ are normalized.

By the triangle inequality,
\beqa
\Trnorm{\Gamma_n - \Gamma} 
&\le& \Trnorm{\Gamma_n - P \Gamma_n P} + \Trnorm{P\Gamma_n P - P \Gamma P}
\nonumber \\
&+& \Trnorm{P\Gamma P - \Gamma}.
\label{triangle1}
\eeqa
Here, $P$ is a finite-dimensional projection, which can be chosen 
(depending on $\Gamma$) to make 
the last term smaller than $\epsilon^2/72$.  By weak convergence,
$\Trnorm{P\Gamma_n P - P \Gamma P} < \epsilon^2/72$ for $n$ greater than
some $M$.  
By Lemma \ref{matrix-blocks}, $n > M$ then implies
$\Trnorm{\Gamma_n - P\Gamma_n P} < \epsilon/2$, 
since $\Trnorm{\Gamma_n - P \Gamma_n P} 
\le \Trnorm{P\Gamma_n P - P \Gamma P}
+ \Trnorm{P\Gamma P - \Gamma}$.
Now plugging into Eq. (\ref{triangle1}) completes the proof.
\end{proof}

The next lemma is not new\cite{enss}, but proofs in the
literature are sketchy.
\begin{lem}
$H_0 + 1$ is locally compact.  That is, if $\Lambda$ is a bounded
region of ${\mathbb R}^{3N}$, and $Q_\Lambda$ denotes projection
onto $\Lambda$, then $Q_\Lambda (H_0+1)^{-1}$ is a compact operator.
\label{H0-lcl-cpt}
\end{lem}
\begin{proof}
First, note that $(1+T+V_{ee})^{-1} = 
(1+T)^{-1} - (1+T)^{-1} V_{ee} (1+T+V_{ee})^{-1}$.
Since
the interaction is relatively bounded with respect to the
kinetic energy, {\it i.e.,} $V_{ee} (1+T+V_{ee})^{-1}$ is bounded,
it suffices to prove this result for $H_0 = T$.

$1+T = (1-\nabla^2) = h^2$, in appropriate units.
The inverse of $h^{2m}$ has integral kernel
\begin{equation}
G_m(x,y) = \int \frac{ e^{ i{\bm k}\cdot({\bm x} -{\bm y}) } }{(1+k^2)^m} \, 
\frac{dk}{(2\pi)^d}.
\nonumber
\end{equation}
$G_m(x,y)$ is bounded by $c |x-y|^{2m-3N}$ as $|x-y| \to 0$, as follows by 
scaling, and for large $|x-y|$, $G_m(x,y)$ falls off exponentially.

Thus, for $m$ large enough,
the Hilbert-Schmidt norm of $Q_\Lambda(x) G_m(x,y)$, given
by $\int_{x\in \Lambda} |G_m(x,y)|^2 \, dx \, dy$ is finite,
so $Q_\Lambda h^{-2m}$ is Hilbert-Schmidt and {\it a fortiori\/} 
compact.  Thus, $h^{2m}$ is locally compact.

Now, if $A \ge 1$ is locally compact, then so is $A^{1/2}$.  
To see this, let $u_n$ be a sequence of vectors converging weakly to zero.
To show that $A^{-1/2}Q u_n \to 0$, note
that $\|A^{-1/2}Q u_n\|^2 = \langle Q u_n | A^{-1} Q u_n\rangle$.  
By assumption, $A^{-1} Q u_n \to 0$, so also $\|A^{-1/2} Q u_n\|^2 \to 0$.
Thus, local compactness of positive operators is preserved by taking
square roots.  
Starting from the locally compact large power $h^{2^n}$, repeated
application shows $h$ also to be locally compact.
\end{proof}

\begin{thm}
\label{tight-cpt}
If $\Delta$ is a tight set of densities, and $M<\infty$, then
${\mathcal S}({\Delta}) \cap {\mathcal E}^{-1}(0,M]$
is relatively compact.
\end{thm}
\begin{proof}
Choose sets $\Lambda_{1/n}$ corresponding to $\Delta$ as described
before Lemma \ref{PDP-cpt}.  
Take $\Gamma \in 
{\mathcal S}({\Delta}) \cap {\mathcal E}^{-1}(0,M]$,
and note that 
$\Gamma - P_{\Lambda_{1/n}} \Gamma P_{\Lambda_{1/n}}$
is the part of the state which puts at least one particle outside
$\Lambda_{1/n}$, so that its norm is less than $1/n$:
$\Trnorm{ \Gamma - P_{\Lambda_{1/n}} \Gamma P_{\Lambda_{1/n}} } < 1/n$.
This shows, via Lemma \ref{PDP-cpt}, that 
${\mathcal S}({\Delta}) \cap {\mathcal E}^{-1}(0,M]$ is
within $1/n$ of the totally bounded set 
$P_{\Lambda_{1/n}} {\mathcal E}^{-1}(0,M] P_{\Lambda_{1/n}}$.
Since $n$ is arbitrary, 
${\mathcal S}({\Delta}) \cap {\mathcal E}^{-1}(0,M]$ is
itself relatively compact.
\end{proof}

With Thm. \ref{tight-cpt} and Lemma \ref{E0-lsc} in hand,
it is now easy to prove the main theorems.
\begin{thm}[{\bf Existence of Levy search solutions}]
\label{minima-nobox}
For every ${\bm \rho}\in {\PD}$, 
${\slbar({\bm \rho})} \not= \emptyset$.
That is, there is a
mixed state $\Gamma \in \slm({\bm\rho})$ such that
$\Tr H_0 \Gamma = \FL({\bm \rho})$.
\end{thm}
\begin{proof}
The set of densities leading to cg-density 
${\bm \rho}$ is a tight set.  Apply Thm. \ref{tight-cpt}
to see that 
$\slm({\bm\rho}) =
\slm({\bm\rho})\cap {\mathcal E}^{-1}(0,{\FL}^{max}]$ is
compact.
Since ${\mathcal E}$ is lower semicontinuous (Lemma \ref{E0-lsc}),
it attains a minimum on $\slm({\bm\rho})$.
\end{proof}

\begin{thm}[{\bf Lower semicontinuity of $\FL$}]
\label{FL-lsc}
$\FL$ is weak lower semicontinuous on ${\PD}$. 
\end{thm}
\begin{proof}
It is required to show that $U_M = \{{\bm\rho}: \FL({\bm\rho}) \le M\}$
is weak closed for any $M$.  Since $U_M$ is convex (because $\FL$ is), 
it suffices to show that $U_M$ is norm closed. 
So, suppose that the sequence $({\bm\rho}_n)_{n=1}^\infty$ is contained
in $U_M$.
Choose $\Gamma_n \in {\slbar}({\bm\rho}_n)$ and apply Thm. \ref{tight-cpt}
with $\Delta = \{{\bm\rho}_1,{\bm\rho}_2,\ldots\}$ 
to see that $\{\Gamma_n\}$ is contained in a compact set, hence contains
a convergent subsequence: $\Gamma_{n_\alpha} \to \Gamma$.
Since the map from states to cg-densities is norm continuous
(each $\hat{\rho}_i$ is a bounded operator),
this means that $\Gamma\in\slm({\bm\rho})$.  By
lower semicontinuity of ${\mathcal E}$ (Lemma \ref{E0-lsc}), 
${\mathcal E}(\Gamma) \le M$.  Thus, $\FL({\bm\rho}) \le M$.  
\end{proof}

The ground state energy for ${\bm v}\in\ell^\infty$ is the infimum
of the spectrum of $H({\bm v})$, also given by
$$
E({\bm v}) = \inf \{ \FL({\bm\rho})+{\bm v}\cdot{\bm\rho}: {\bm\rho}\in\ell^1\}.
$$
It can happen that there is no state which actually achieves the infimum.
This definition actually makes sense for ${\bm v} \in {\mathcal V}$.
Also, ${\bm v}$ is bounded below in terms of $E({\bm v})$:
\begin{equation}
\inf_i v_{i} \ge E({\bm v}) - {\FL}^{max}.
\label{v-lower-bd}
\end{equation}

Since $\FL$ is clearly convex, and weak lower semicontinuous by
Thm. \ref{FL-lsc},
it follows (ET: Prop. I.4.1, vT: cor. to lemma 6.12)
that $\FL$ (on $\ell^1$) and $E$ (on $\ell^\infty$) are 
a Legendre-Fenchel transform pair, up to signs.
That is,
\beq
\FL({\bm \rho}) = 
\sup \{ E({\bm v})-{\bm v}\cdot{\bm\rho}: {\bm v}\in\ell^\infty\}.
\label{FL-as-LF-xform}
\eeq
Again, as in the confined case, ${\bm \rho}$ is cg-EV-representable if 
and only if there is ${\bm v}\in{\mathcal V}$ such that
such that $\FL({\bm \rho}_0) + \int v \rho_0 \, dx \le
\FL({\bm \rho}) + \int v \rho \, dx$, for all ${\bm \rho}$.  
Note that this is not the same thing as saying that the
supremum on the right-hand side of Eq. (\ref{FL-as-LF-xform})
is attained.  That would mean that the cg-density in question
was a ground state cg-density of a potential in 
$\ell^\infty$; not all are so.

It is this last observation that requires the method of proof
of cg-EV-representability to differ substantially from that
employed in Section \ref{box}.
In a remarkable paper, Chayes, Chayes and Ruskai\cite{ccr}
investigated the representability problem on a lattice and showed 
that all densities are EV-representable.  
Surprisingly, their method of proof can be adapted to the current 
situation with only minor modifications. 
The general idea is that the potential which represents $\bm\rho$ 
should be the one which maximizes $E({\bm v}) - {\bm \rho}\cdot{\bm v}$.  
Find a maximizing sequence ${\bm v}^\alpha$, and, using a diagonal 
construction and the countability of the set of cells, extract a 
subsequence which converges 
on each cell, to a candidate potential ${\bm v}$.  
Cell-wise convergence is too vague to ensure much about $H({\bm v})$, 
so the second part of the proof
establishes that $H({\bm v}^\alpha)$ actually converges to $H({\bm v})$
in strong resolvent sense, which is enough to show that it has a
ground state in $\slm({\bm\rho})$. 
 
Specific results from Reed and 
Simon\cite{RS} or Weidmann\cite{weidmann} are indicated by RS or W.
\begin{thm}
If ${\bm \rho} \in \P0$, then  
${\bm \rho}$ is cg-EV-representable by a potential in ${\mathcal V}$.
\label{v-rep-nobox}
\end{thm}
\begin{proof}
A constant can always be added to ${\bm v}$ to make $E({\bm v})=0$,
so
\beq
\FL({\bm \rho}) = \sup\{ -{\bm v}\cdot{\bm \rho} \, :\, E({\bm v}) = 0 \}.
\nonumber
\eeq
Let ${\bm v}^\alpha$ be a maximizing sequence for the right-hand side.
That is,
\beq
\int_{v^\alpha < 0} \rho |v^\alpha| \, dx
- \int_{v^\alpha > 0} \rho |v^\alpha| \, dx
\nearrow \FL({\bm \rho}).
\label{max-seq}
\eeq
Since $E({\bm v}^\alpha) = 0$, ${v}^\alpha < 0$ somewhere.
Define, for each $\alpha$, a cg-density 
\beq
\rho^\alpha = 
\frac{\Nt \theta(-v^\alpha) \rho}{
\int \theta(-{v^\alpha}) \rho\, dx} 
\ge \theta(-v^\alpha) \rho.
\nonumber
\eeq
Then,
\beq
\FL^{max} \ge \FL({\bm \rho}^\alpha) 
\ge \int_{v^\alpha < 0} \rho^\alpha |v^\alpha| \, dx
\ge \int_{v^\alpha < 0} \rho |v^\alpha| \, dx.
\nonumber
\eeq
So,
$\int_{{v^\alpha}<0} \rho |v^\alpha| \, dx$ is bounded, uniformly in
$\alpha$, hence, by Eq. (\ref{max-seq}),
$\int_{{v^\alpha}>0} \rho |v^\alpha| \, dx$ is, too. 
Thus, $\int \rho |v^\alpha| \, dx$ is bounded by some constant, and
\beq
|v_i^\alpha| \le \frac{c}{\rho_i \Omega}.
\label{va-bd}
\eeq
Since each $v_i^\alpha$ thus lies in a bounded interval, a diagonal
construction will yield a subsequence
${\tilde{\bm v}^\alpha}$ which converges on each cell, though
not in general with any uniformity. 
By relabeling, we can just write $v_i^\alpha$ for this subsequence,
so $v_i^\alpha \to v_i$ for each $i$.
This defines our candidate potential ${\bm v}$ to have $\bm \rho$
as ground state cg-density.  

$H({\bm v})$ makes sense because $v_i$ is everywhere
bounded below by Eq. (\ref{v-lower-bd}), and everywhere finite by
Eq. (\ref{va-bd}), so ${\bm v} \in{\mathcal V}$.
Let ${\mathcal D}$ be the set of smooth pure states with density nonzero only 
in a finite number of cells; since the Coulomb repulsion is positive and
locally $L^2$, ${\mathcal D}$ is a common domain of essential self-adjointness
for all $H({\bm v}^\prime)$ with ${\bm v}^\prime\in{\mathcal V}$  
(RS: Thm. X.28, Hislop and Sigal: Thm. 8.14).
For $\phi$ a unit vector in ${\mathcal D}$,
$0 \le \langle \phi| H({\bm v}^\alpha) \phi \rangle$ since
$E({\bm v}^\alpha) = 0$, so 
\beq
0 \le \lim_{\alpha \to \infty} \langle \phi| H({\bm v}^\alpha) \phi \rangle
= \langle \phi| H({\bm v}) \phi \rangle < \infty,
\nonumber
\eeq
because the convergence of the potential is uniform on the support of $\phi$.
Thus, since ${\mathcal D}$ is a core, $H({\bm v}) \ge 0$.
Further, $\| [H({\bm v})-H({\bm v}^\alpha)]\phi \| = \| [v^\alpha -v]\phi \|
\to 0$, which shows convergence of $H({\bm v}^\alpha)$ to
$H({\bm v})$ on $\mathcal D$.  This suffices (RS:VIII.25 or W:Thm. 9.16) 
to establish strong resolvent convergence.

Now, find $\Gamma_0 \in {\slbar({\bm \rho})}$, so that
$$
{\mathcal E}(\Gamma_0) = \FL({\bm \rho}) = 
\lim_{\alpha\to\infty} \int -\rho v^\alpha \, dx.
$$
A $\Gamma_0$ which satisfies the first equality exists by 
Theorem \ref{minima-nobox}, and the second equality holds by Eq. (\ref{max-seq}).
Since $e^{-x}$ is a bounded function on $x > 0$,
$e^{-H({\bm v}^\alpha)} \to e^{-H({\bm v})}$ in strong operator topology
(RS:VIII.20, or W:Thm. 9.17).
Then $H({\bm v}) \ge 0$ implies that
\beqa
1 &\ge& \Tr \Gamma_0 e^{-H({\bm v})}
= \lim_{\alpha \to \infty} \Tr \Gamma_0 e^{-H({\bm v}^\alpha)}
\nonumber
\\
&\ge& e^{-{\mathcal E}(\Gamma_0)}
\lim_{\alpha \to \infty}
\exp \left[ -\int \rho v^\alpha \, dx \right]
= 1,
\label{jensen-squeeze}
\eeqa
by use of Jensen's 
inequality and the fact that 
$({\bm v}^\alpha)$ is a maximizing sequence for $\FL({\bm \rho})$.
Thus, $\Tr \Gamma_0 e^{-H({\bm v})} = 1$, which, since 
$H({\bm v}) \ge 0$, shows that $\Gamma_0$ is concentrated on
the ground state manifold of 
$H({\bm v})$.
\end{proof}

\section{indefinite particle number}
\label{fock}

Open systems\cite{pplb} are also of interest for density
functional theory.  In this case, the total particle number
can fluctuate -- the system is in equilibrium with a
particle reservoir at some chemical potential, which we
take to be zero without loss of generality.
The theory of the previous two sections extends in full
to this case of indefinite particle number.  We will explicitly
discuss only the unconfined case.

The relevant state space is the Fock space
\beq
{\mathcal F} = \bigoplus_{m=0}^\infty {\mathcal H}_{m},
\nonumber
\eeq
where ${\mathcal H}_{m}$ denotes the $m$-particle Hilbert space.
The set of states which interest us is not the entirety of
${\mathcal L}_1^{+,1}({\mathcal F})$, but the subset $\slm$
of states which are diagonal in particle number, {\it i.e.,}
those of the form $\sum_{m=0}^\infty a_m \Gamma^{(m)}$,
where $\Gamma^{(m)}$ is a normalized $m$-particle mixed state,
and $\sum a_m = 1$.  Below, we write $\Gamma^{(m)}$ for the
normalized $m$-particle part of any mixed state $\Gamma$.
The following are natural generalizations of previous definitions
to the Fock-space setting.
The subset of states in $\slm$ which have density ${\bm \rho}$ is
denoted $\slm({\bm\rho})$, and the grand canonical density functional
is defined
\beq
F({\bm \rho}) \defeq \inf \left\{
{\Tr} \Gamma H_0 : \, \Gamma \in {\slm}({\bm \rho}) \right\},
\label{Fgc}
\eeq
where $H_0$ is simply the direct sum of the kinetic-plus-interaction
Hamiltonians for all particle numbers.
Notice that the mixed states in Eq. (\ref{Fgc}) are not restricted
to two consequtive total particle numbers, in contrast to the
procedure introduced by Perdew, Parr, Levy and Balduz\cite{pplb}.
The ground state energy for potential $\bm v$ is the infimum
of the spectrum of $E({\bm v})$ and a state which realizes it is
a ground state.  As before, we can also write
\beq
E({\bm v}) 
\defeq \inf \left\{ 
F({\bm \rho}) +{\bm v}\cdot{\bm\rho} \, : \, {\bm\rho}\in \ell^1 \right\},
\nonumber
\eeq
and a ground state is a state which realizes the infimum.
Now, the additive constant is important.  Decreasing it is likely to
result in a ground state with more particles.  For some potentials,
even though bounded below, $E({\bm v}) = -\infty$ because the energy
can be lowered indefinitely by adding particles.

The proofs of Lemma \ref{E0-lsc} and of the Hohenberg-Kohn Theorem 
apply to the current situation with almost no essential changes.
The density which is everywhere zero is the exception.  It is
certainly EV-representable, but not uniquely.

We now sketch the required alterations to the proofs of 
Sec. \ref{nobox}.  
Consider generalizing Thm. \ref{tight-cpt} to a tight
sets of densities $\Delta$ all with the same total particle number 
$\Nt$.  ${\mathcal S}(\Delta)$ now refers to the corresponding
set of $\Nt$-diagonal Fock-space mixed states.
Choose $\epsilon > 0$.
If $\Tr \Gamma \hat{N}_{tot} =N$,
then clearly
$\sum_{m > N/\epsilon} \alpha_m < \epsilon$.
Similarly, if ${\mathcal E}(\Gamma) \le M$, then
$\sum_{m \in I} \alpha_m < \epsilon$, where
\beq
I = \{ m\in {\Bbb N}: {\mathcal E}(\Gamma^{(m)}) > M/\epsilon\}.
\nonumber
\eeq
According to Thm. \ref{tight-cpt}, the set $A_m$ of $m$-particle
mixed states with energy less than or equal to $M/\epsilon$ and
density less than or equal to some density in $\Delta$ is
compact.  Since ${\mathcal S}(\Delta)$ is within $2\epsilon$ of
the convex hull of $\cup_{m\le N/\epsilon} A_m$,
and $\epsilon$ is arbitrary, 
${\mathcal S}(\Delta)\cap {\mathcal E}^{-1}(0,M]$ is itself totally 
bounded, {\it i.e.,} relatively compact.

With Lemma \ref{E0-lsc} and Thm. \ref{tight-cpt} thus extended
to Fock space, the proofs that $\underline{\mathcal S}({\bm\rho}) \not= \emptyset$
and that $F$ is lower semicontinuous proceed just as in Thms. \ref{minima-nobox}
and \ref{FL-lsc}.
Thus, we can write, as before,
\beq
\FL({\bm \rho}) = 
\sup \{ E({\bm v})-{\bm v}\cdot{\bm\rho}: {\bm v}\in\ell^\infty\}.
\label{F-LF}
\eeq
What is different about the situation now is that the effective 
domain of $\FL$ is all cg-densities: positivity is still required,
but normalization is not.

The final stage is to adapt the proof of Thm. \ref{v-rep-nobox} to
show that any cg-density, everywhere strictly positive, is
Fock-space cg-EV-representable.
The trick of adjusting the additive constant in the potential to obtain
$E({\bm v})=0$ had a significant part in the proof of Thm. \ref{v-rep-nobox},
where it was both a labor-saving device and a means of assuring that
the potential did not drift off to infinity.  
In the Fock space setting, it is neither possible nor needed, and
we proceed as follows.

Given ${\bm\rho}$, find a sequence of potentials ${\bm v}^\alpha$ such that
\beq
E({\bm v}^\alpha) -{\bm v}^\alpha \cdot{\bm \rho} \, \nearrow F({\bm \rho}).
\nonumber
\eeq
Since $F({\bm\rho}) > 0$,
for $\alpha$ large enough
\beq
E({\bm v}^\alpha) > {\bm v}^\alpha \cdot{\bm \rho} \ge
{\bm v}_{min}^\alpha \Nt,
\label{Ebnd-by-vmin}
\eeq
where ${\bm v}_{min}^\alpha = \inf_i {\bm v}_i^\alpha$.
But also, by considering a state with any number $M$ of particles 
all located in a cell where the potential is favorable,
$E({\bm v}^\alpha) \le c^\prime M^3 + {\bm v}_{min}^\alpha M$.
Combining these inequalities,
\beq
{\bm v}_{min}^\alpha \ge -\frac{c^\prime M^3}{M-\Nt}, 
\quad \mbox{\rm if }M > \Nt.
\nonumber
\eeq
Choosing $M$ to be the next integer larger than $\Nt$, for example,
this shows that ${\bm v}_{min}^\alpha$ is bounded below independently of
$\alpha$ for $\alpha$ large enough.  And, inserting this bound back
into inequality (\ref{Ebnd-by-vmin}) yields a lower bound for
the $E({\bm v}^\alpha)$ as well.
Thus, by a diagonal argument as before, a subsequence
(written ${\bm v}^\alpha$ without loss) exists for which
$v_i^\alpha$ converges to $v_i$ for each $i$, and
$E({\bm v}^\alpha)$ to $E_\infty$ as $\alpha\to\infty$.

The rest of the proof proceeds largely as before, with a few
$E_\infty$ inserted.  The appropriate domain of essential 
self-adjointness ${\mathcal D}$ is now the set of 
states with density supported in a finite number of cells, and
also with components in only a finite number of ${\mathcal H}_m$.
The inequality showing positivity of $H({\bm v})$ becomes
\beq
E_\infty \le \langle \phi| H({\bm v}) \phi \rangle < \infty.
\nonumber
\eeq
Since $\{E({\bm v}^1),E({\bm v}^2),\ldots,E_\infty\}$ is
bounded below,
$\{ H({\bm v}^1),H({\bm v}^2),\ldots,H({\bm v})\}$ is also,
so
$e^{-H({\bm v}^\alpha)} \to e^{-H({\bm v})}$ in strong operator topology.
With $\Gamma_0 \in {\slbar({\bm \rho})}$, as before,
the counterpart of Eq. (\ref{jensen-squeeze}) is
\beq
e^{-E_\infty} \ge \Tr \Gamma_0 e^{-H({\bm v})}
= \cdots = e^{-E_\infty}.
\nonumber
\eeq

\section{conclusion}
\label{conclusion}

At the simplest level, the results presented in this 
paper merely increase our knowledge about the 
V-representability problem.  Within every coarse-grained 
equivalence class of densities, there is at least one which 
is EV-representable.  In addition, there is one such which 
is representable by a unique potential which is constant on 
cells, and bounded below, though not necessarily 
(in the unconfined case) bounded above.  
Extensions of these results to spin density functional
theory and multicomponent systems is straightforward.

Valuable as those observations may be, that way of putting it
misses the bigger idea that a coarse-grained point of view may 
be a generally superior way to look at the foundations of density 
functional theory.  From that perspective, what has been achieved
is support for the idea in that EV-representability is shown
to not be a problem {\em at all} in a coarse-grained framework.
Little is lost by this change, since the coarse-graining scale can 
be chosen as the size of an atomic nucleus, for example, 
and most knowledge we may have of the fine-grained level can be
incorporated relatively easily.  

Establishing EV-representability is the first 
step toward controlling the local structure of the Lieb 
functional.  Because, if $\FL$ is differentiable at $\rho$, 
in any sense, the derivative is
given by the representing potential.  With regard to this
question, the coarse-grained theory has taken one step.
We have shown that $\FL$ is differentiable in the confined
case, by taking advantage of the finite-dimensionality.  
The most immediate outstanding problem at this point is to demonstrate
some sort of differentiability in the unconfined situation.
General principles avail us nothing because $X_N^+$ has empty 
interior relative to $X_N$.  (The fine-grained theory also has 
this problem\cite{unpub}. A widespread view that $\FL$ is 
G\^ateaux differentiable at EV-representable densities is
in error.) 
Hopefully, it will be possible to show that if 
$\eta_i \not = 0$ for only {\em finitely many\/} $i$ and 
$\sum \eta_i = 0$, then $\FL({\bm\rho}+s{\bm \eta})$ 
is differentiable as a function of $s$ at $s=0$.
This property is weaker than G\^ateaux differentiability, 
but probably not in any important way.

\medskip
\noindent{\bf Acknowledgements.}
The author thanks Vin Crespi and Jorge Sofo for advice, and
the Penn State Center for Nanoscale Science for financial support.

\copyright 2006 American Institute of Physics

\end{document}